\documentstyle[aps,prl,epsf]{revtex}
\def\narrowtext{} \tighten \twocolumn
\input epsf.sty
\begin{document}
\draft

\title{Quasiparticle Formation and Optical Sum Rule Violation in Cuprate
Superconductors}
\author{M. R. Norman$^{1,2}$ and C. P\'{e}pin$^2$}
\address{$^1$Materials Science Division, Argonne National Laboratory, 
Argonne, IL 60439, USA}
\address{$^2$SPhT, L'Orme des Merisiers, CEA-Saclay, 91191 Gif-sur-Yvette,
France}
\address{%
\begin{minipage}[t]{6.0in}
\begin{abstract}
Using a simple model for the frequency dependent scattering rate, we
evaluate the in-plane optical integral for cuprate superconductors in the
normal and superconducting states.  In the overdoped region,
this integral is conserved.
In the optimal and underdoped region, though,
the optical integrals differ, implying a lowering of
the in-plane kinetic energy in the superconducting state.
This sum rule violation, due to the difference
of the non Fermi liquid normal state and the superconducting Fermi
liquid state, has a magnitude comparable to recent experimental results.
\typeout{polish abstract}    
\end{abstract}
\end{minipage}}
\maketitle
\narrowtext

In superconductors, there is a dramatic change in the 
conductivity due to opening of an excitation gap in the finite frequency
response, and the formation of a zero
frequency $\delta$ function peak representing the dissipationless response of
the condensate.  The change is such as to preserve
the optical sum rule, in that the finite frequency weight removed
by the opening of the excitation gap is recovered by the condensate
peak\cite{TFG}.

In cuprate superconductors, though, there is experimental evidence that
the sum rule is violated for c-axis conductivity\cite{BASOV}.  Over the
measured frequency range, more weight is present in the condensate
peak than can be accounted for by the loss of finite frequency weight.
Since the total optical sum rule must be preserved, the
extra weight in the condensate peak is coming from outside this
frequency range.  This is unusual, since in classic
superconductors, the change in the optical integral is exhausted
over a frequency range of order 4$\Delta$, where $\Delta$ is the superconducting
gap.  Anderson\cite{PHIL} has stressed that such
sum rule violations are connected to the lack of quasiparticle poles in the
normal state, and their emergence in the superconducting state.
This is unlike the premise of BCS theory, where quasiparticles are assumed
to exist in the normal state.

Although the sum rule violation for the c-axis response is profound, its
contribution to the condensation energy is small due to the smallness of the 
c-axis kinetic energy in the cuprates.  If kinetic energy effects are to play 
a role in the condensation energy, then they must be coming from the in-plane 
response, since the in-plane kinetic energy is quite large, of order an 
eV\cite{HM1}.  This largeness, though, means that the violation is difficult to 
see.  That is, a 1 meV change in
the kinetic energy would represent $\sim$1\% change in the optical integral.
Recently, though, two groups have claimed to observe such a change.
Ellipsometry
data on optimal and underdoped Bi2212\cite{VDM} have been quantified
as corresponding to a change in the kinetic energy of 1 meV.
The same kinetic energy change has been inferred from reflectance
data\cite{NICOLE} on an underdoped Bi2212 film,
%MRN - addition
though no such change could be resolved in an overdoped film.
These results are intriguing, since a 1 meV kinetic energy
savings per plane is in excess of the condensation energy inferred
from specific heat data\cite{LORAM}.

In this paper, using a simple model for
the frequency dependent scattering rate based on angle resolved
photoemission (ARPES) and infrared data, we calculate the change in
the optical integral from the normal to the superconducting state, and find
its sign and magnitude to be comparable to these recent findings.

The full optical integral, integrating over all energy bands,
is proportional to the bare carrier density over the bare electron mass, and
thus must be conserved.  Of greater interest here is the optical response
of the band around the Fermi energy, correlating with the experimental
data which are typically integrated out to an energy of order the
plasma frequency (1 eV).  This leads to a consideration of the
single band sum rule\cite{KUBO}
\begin{equation}
\int_0^{\infty} Re \sigma_{xx}(\omega) d\omega =
\frac{\pi e^2 a^2}{2 \hbar^2 V} E_K
\end{equation}
where the restriction of $\sigma$ to the single band response is implicit, and
where $a$ is the in-plane lattice constant, $V$ the unit cell volume, and
\begin{equation}
E_K=\frac{2}{a^2 N}
\sum_k \frac{\partial^2 \epsilon_k}{\partial k_x^2} n_k
\end{equation}
with $N$ the number of k vectors, $\epsilon_k$ the bare dispersion as defined
by the effective single band Hamiltonian\cite{COND}, and $n_k$ the momentum
distribution function.  For a Hamiltonian with near neighbor hopping\cite{HM1},
$E_K$ is equivalent to minus the kinetic energy
%MRN -addition 
($E_{kin} \equiv \frac{2}{N} \sum_k \epsilon_k n_k$),
but in general these two quantities differ.

For free electrons, the inverse mass tensor is a constant in
momentum, and thus
this integral is conserved due to charge conservation.  This is not generally 
the case, since the sum of the inverse mass
tensor over the Brillouin zone vanishes\cite{AM}.  When considering the
change in this integral between different electronic states, the emphasis
in the past has been on a possible change in the inverse mass
tensor\cite{HM1,KLEIN}.  In general, though, we expect $\epsilon_k$ to be
invariant, and therefore the change should instead be due to changes in
$n_k$.
A simple case is BCS theory\cite{TFG}, where the kinetic
energy increases in the superconducting state 
due to particle-hole mixing.  If a near neighbor tight binding model applied, 
the BCS
optical integral would be smaller in the superconducting state than in the
normal state, opposite to the recent experimental results.

The BCS model, though, assumes the existence of quasiparticle poles
in the normal state.  It is straightforward to demonstrate that the kinetic
energy can
indeed be lowered in the superconducting state if the normal state is a non
Fermi liquid and the superconducting state a Fermi liquid\cite{COND}.  This
occurs if the effect of quasiparticle formation on sharpening
$n_k$ is larger than the smearing due to
particle-hole mixing.  This effect
is anisotropic in momentum, due to anisotropies in the scattering
rate and
the d-wave order parameter.  Given these anisotropies and the
anisotropy of the mass tensor, it is not obvious what the effect of the
kinetic energy lowering will be on the optical integral, since a near
neighbor tight binding model is inadequate to describe $\epsilon_{\bf k}$.
In addition, ARPES
measurements indicate a substantial doping dependence of the scattering rate,
which implies that the sum rule violation will also be doping dependent.

We start by considering a simple model for the
frequency dependent scattering rate, based on fits to ARPES data at the
$(\pi,0)$ point\cite{ND}.  This was used in work on the
condensation energy\cite{COND} and
the c-axis sum rule\cite{IM1}.  The model assumes a large frequency independent
scattering rate in the normal state, consistent with the broad Lorentzian
lineshapes.
In the superconducting state, the broad peak is replaced by a sharp
peak at the superconducting gap energy, followed at higher binding
energy by a spectral dip, then a broad maximum (the ``hump").
This change
is modeled by cutting off Im$\Sigma$ at the energy of the spectral
dip.  The resulting $\Sigma$ is
\begin{equation}
\Sigma_{\Gamma} = \frac{\Gamma}{\pi}
{\rm ln}\left|\frac{\omega-\omega_0}{\omega+\omega_0}\right|
-i\Gamma\Theta(|\omega|-\omega_0)
\end{equation}
where $\omega_0$ is the spectral dip energy.
This self-energy is then used in the spectral function\cite{SCHRIEFFER}
\begin{equation}
A = \frac{1}{\pi}{\rm Im}
\frac{Z\omega+\epsilon}{Z^2(\omega^2-\Delta^2)-\epsilon^2}
\end{equation}
where $Z = 1-\Sigma/\omega$.  For this form of $\Sigma$, the spectral function
has two $\delta$
functions located at $\pm E$, where
$E$ satisfies the pole condition (denominator of Eq.~4 vanishes).
Such poles
always exist for $E < \omega_0$ because of the log divergence of Re$\Sigma$
at $\pm\omega_0$.  The weight of the poles are
determined as \cite{MAHAN} $|dA^{-1}(\pm E)/d\omega|$.
In addition, there
are incoherent pieces for $|\omega| > \omega_0$.

For now, we assume $\Gamma$ is $k$ independent.  $\omega_0$ is also
assumed to be $k$ independent, as implied by ARPES experiments\cite{ADAM}.
$\epsilon_k$ is taken from a six parameter tight binding fit to normal state
ARPES data\cite{NORM95}.
For the order parameter, the d-wave form $\cos(k_xa)-\cos(k_ya)$ is
assumed.  The $k$ sum is done using a 100 by 100 grid in the irreducible
quadrant of the zone.  The quasiparticle
pole weight contribution to $n_k$ is analytic\cite{LOR}.  The incoherent 
contribution is evaluated by trapezoidal integration.
We consider the T=0 limit, and thus $n_k=\int_{-\infty}^0 A(\omega) d\omega$.
In practice, the lower cut-off is taken to be -10 eV.  In the
normal state with no lower cut-off,
$n_k = 1/2 - \tan^{-1}\left(\epsilon/\Gamma\right)/\pi$.

\begin{figure}
\centerline{
\epsfxsize=0.24\textwidth{\epsfbox{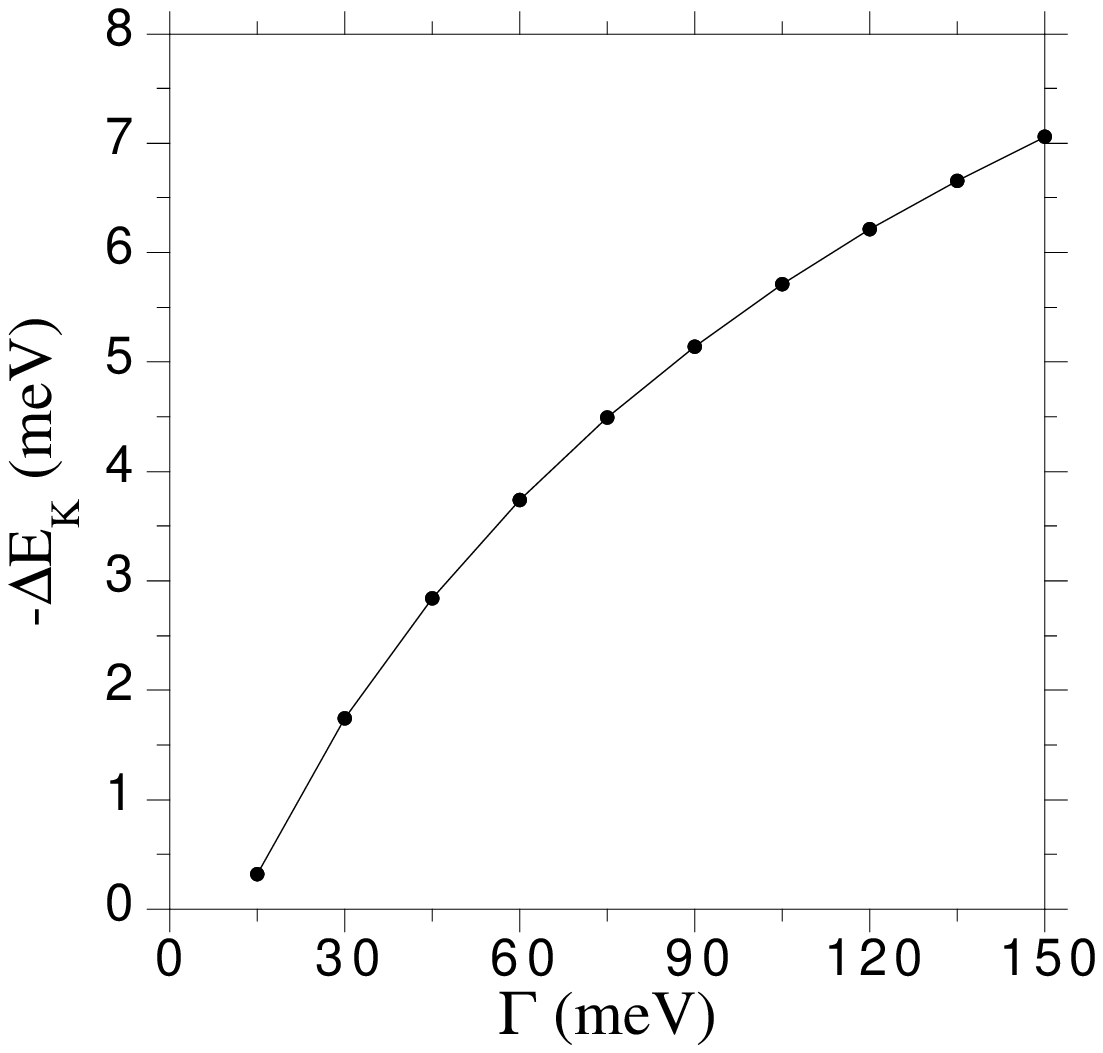}}
\epsfxsize=0.24\textwidth{\epsfbox{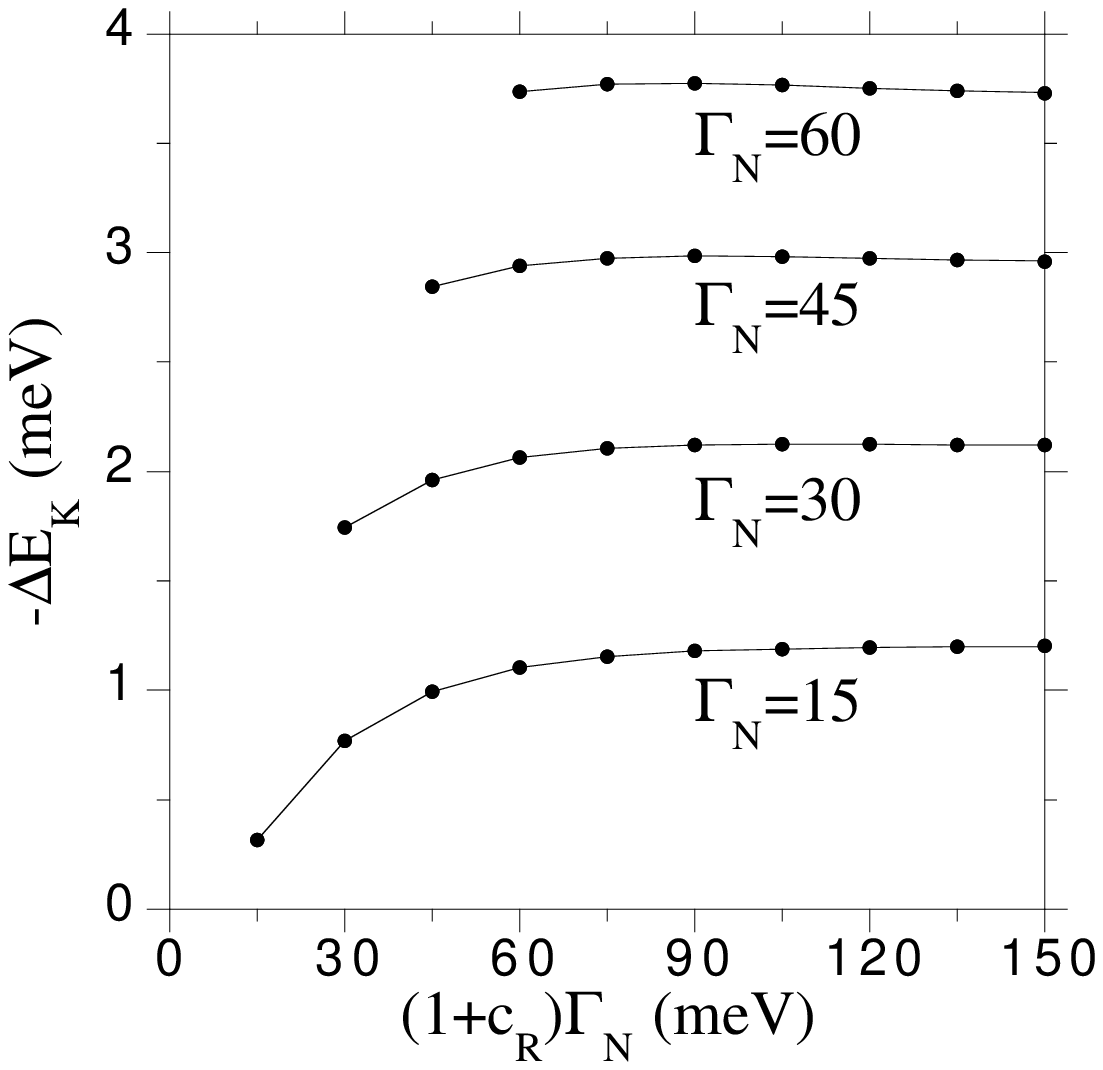}}
}
\vspace{0.3cm}
\caption{$-\Delta E_K$ versus (a) isotropic scattering rate and
(b) antinodal scattering rate (for various nodal scattering rates).
$\omega_0$=71meV and $\Delta_{max}$=32meV.}
\label{fig1}
\end{figure}

In Fig.~1a, $\Delta E_K \equiv E_K^N-E_K^S$ (where $N$ denotes normal state
and $S$ superconducting state) is plotted as a
function of $\Gamma$.  The strong dependence on $\Gamma$
is expected, since as $\Gamma$ increases, the change in $n_k$
($\Delta n_k \equiv n_k^N-n_k^S$)  becomes
increasingly pronounced, leading to a larger sum rule violation.
This implies that the sum rule violation becomes larger as the
doping decreases, since $\Gamma$ from ARPES measurements increases
with underdoping.

One issue with Fig.~1a is the rather large value of the sum rule violation
for realistic values of $\Gamma$
%MRN - addition 
(the antinodal scattering rate from ARPES is $\sim$100 meV for optimal doping).
It is known, though,
that the scattering rate from ARPES is a strong function of momentum\cite{JS1}.
We consider a simple model
for the anisotropy with
$\Gamma_k = \Gamma_N [1 + c_R (\cos(k_xa)-\cos(k_ya))^2/4]$,
where $\Gamma_N$ is the nodal scattering rate and $\Gamma_N (1+c_R)$ the
antinodal one.  In Fig.~1b, we plot $\Delta E_K$ versus $c_R$ for
various $\Gamma_N$, and find that it rapidly saturates with
$c_R$, and thus with the antinodal scattering rate.  We have also
considered the influence of the anisotropic pseudogap on the normal 
state\cite{PHEN}, and found this had little effect on $\Delta E_K$.

To gain further insight, we plot in Fig.~2a the integrand
of $\Delta E_K$ as a function of momentum.  Note that the overall integral is
negative, with negative regions corresponding to unoccupied states near the
d-wave node ($(0,0)-(\pi,\pi)$ Fermi crossing) and occupied states near the
$(\pi/2,0)$ points, and positive
regions to occupied states near the node and unoccupied states near the 
antinode ($(\pi,0)-(\pi,\pi)$ Fermi crossing).  To understand this, we plot
two curves on Fig.~2a, one the Fermi surface, the other the zero of
the inverse mass tensor.  In our model $n_k$ is equal to 1/2 on the Fermi 
surface, and
thus $\Delta n_k$ changes sign there.  Therefore, the
optical integrand, which is the product of the inverse mass tensor times
$\Delta n_k$, changes sign each time one of these two curves is crossed.
From this, one can easily understand the various sign regions in the plot.

\begin{figure}
\centerline{
\epsfxsize=0.20\textwidth{\epsfbox{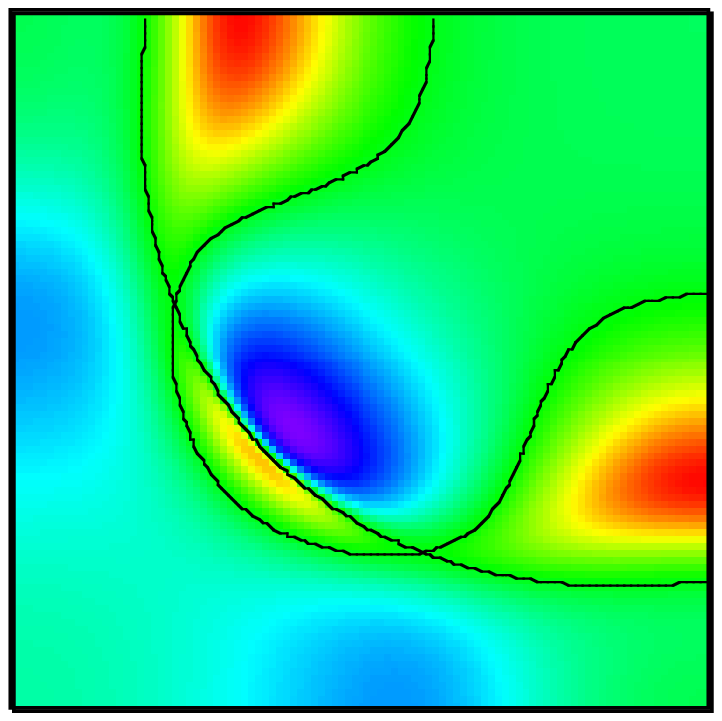}}
\epsfxsize=0.20\textwidth{\epsfbox{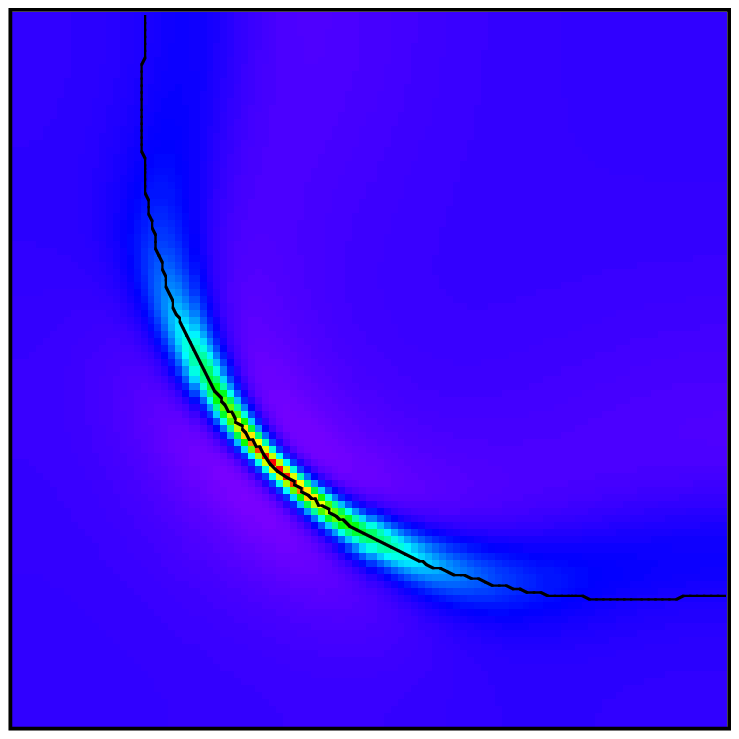}}
}
\vspace{0.3cm}
\caption{(a) $\Delta E_K$ versus $k$ (red is positive, green near zero,
blue negative).  The curves are the Fermi surface and zero of the inverse
mass tensor.
(b) $\Delta(\nabla n_k \cdot \nabla \epsilon_k)$ versus $k$ (red is positive,
blue near zero).  $\Gamma$=150meV, $\omega_0$=71meV, and $\Delta_{max}$=32meV.} 
\label{fig2}
\end{figure}

Perhaps more instructive is to convert the optical integral to the equivalent
one involving $-\nabla n_k \cdot \nabla \epsilon_k$ (using
Greens theorem for periodic functions\cite{AM}).
The resulting integrand is plotted in Fig.~2b, and
as expected, is localized about the Fermi surface.  The important point
is that the integrand peaks at the node.  This can be easily
understood.  In the superconducting state there are quasiparticle poles,
but at the node, $\Delta_k=0$, so there is a true step discontinuity in
$n_k$ there.  As one moves away from the node, $\Delta_k$ increases from zero,
and so $|\nabla n_k|$ decreases in magnitude.
From Fig.~2b, it is easy to appreciate the result of Fig.~1 that the
optical integral is sensitive to the nodal scattering rate and not so sensitive
to the antinodal one.  As our model was
motivated by fitting ARPES data in the antinodal region of the zone, this
indicates that a model based directly on the nodal region
should be considered.

Normal state ARPES
data\cite{JS1} are consistent with a scattering rate of the form
-Im$\Sigma = \Gamma_k + \alpha|\omega|$,
where $\Gamma_k$ has the anisotropy described above
and $\alpha$ is momentum independent\cite{AV}.  For simplicity, we
will assume that both of these terms have an infrared cut-off at
$\omega_0$ as we did for the $\Gamma$ model.
When determining
Re$\Sigma$, it is important to provide an ultraviolet cut-off to Im$\Sigma$.
A hard cut-off at $\omega_c$ leads to a log singularity in
Re$\Sigma$ at $\omega_c$.  Rather, we take Im$\Sigma$ to saturate at
$\omega_c$.  This gives
\begin{equation}
{\rm Re} \Sigma_{\alpha} =
\frac{\alpha}{\pi}\left(\omega
{\rm ln}\left|\frac{\omega^2-\omega_0^2}{\omega^2-\omega_c^2}\right|
+ \omega_c
{\rm ln}\left|\frac{\omega-\omega_c}{\omega+\omega_c}\right|\right)
\end{equation}
where $\Sigma = \Sigma_{\alpha} + \Sigma_{\Gamma}$.
The normal state $\Sigma$ is obtained by setting $\omega_0=0$.
$\omega_0$
is the energy of the dispersion kink along the zone diagonal, which
is the same energy as the spectral dip at $(\pi,0)$\cite{ADAM}, and thus
$\omega_0$ is $k$ independent.
For $\omega_c$, fits to ARPES are consistent with a value of
500 meV\cite{ADAM2}.

\begin{figure}
\centerline{
\epsfxsize=0.24\textwidth{\epsfbox{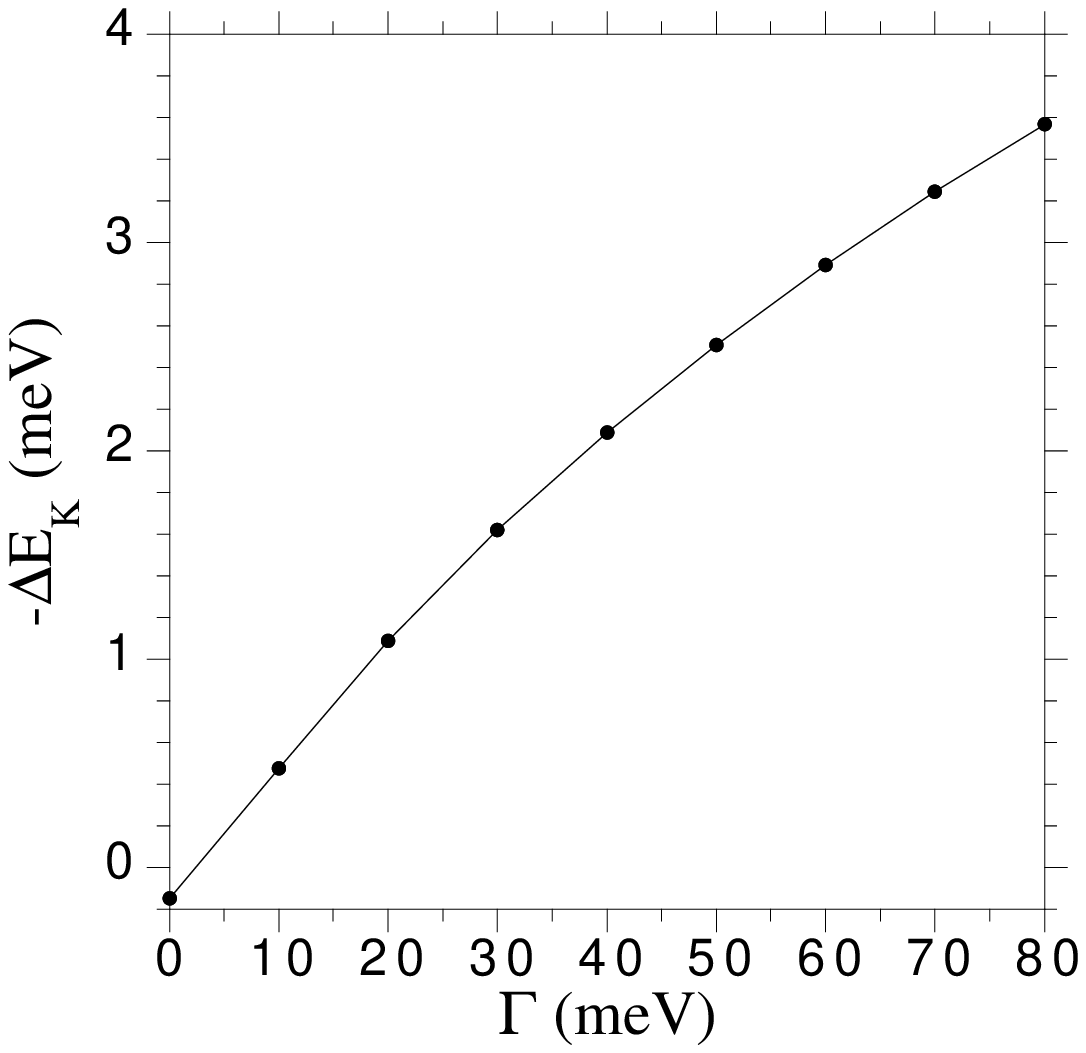}}
\epsfxsize=0.24\textwidth{\epsfbox{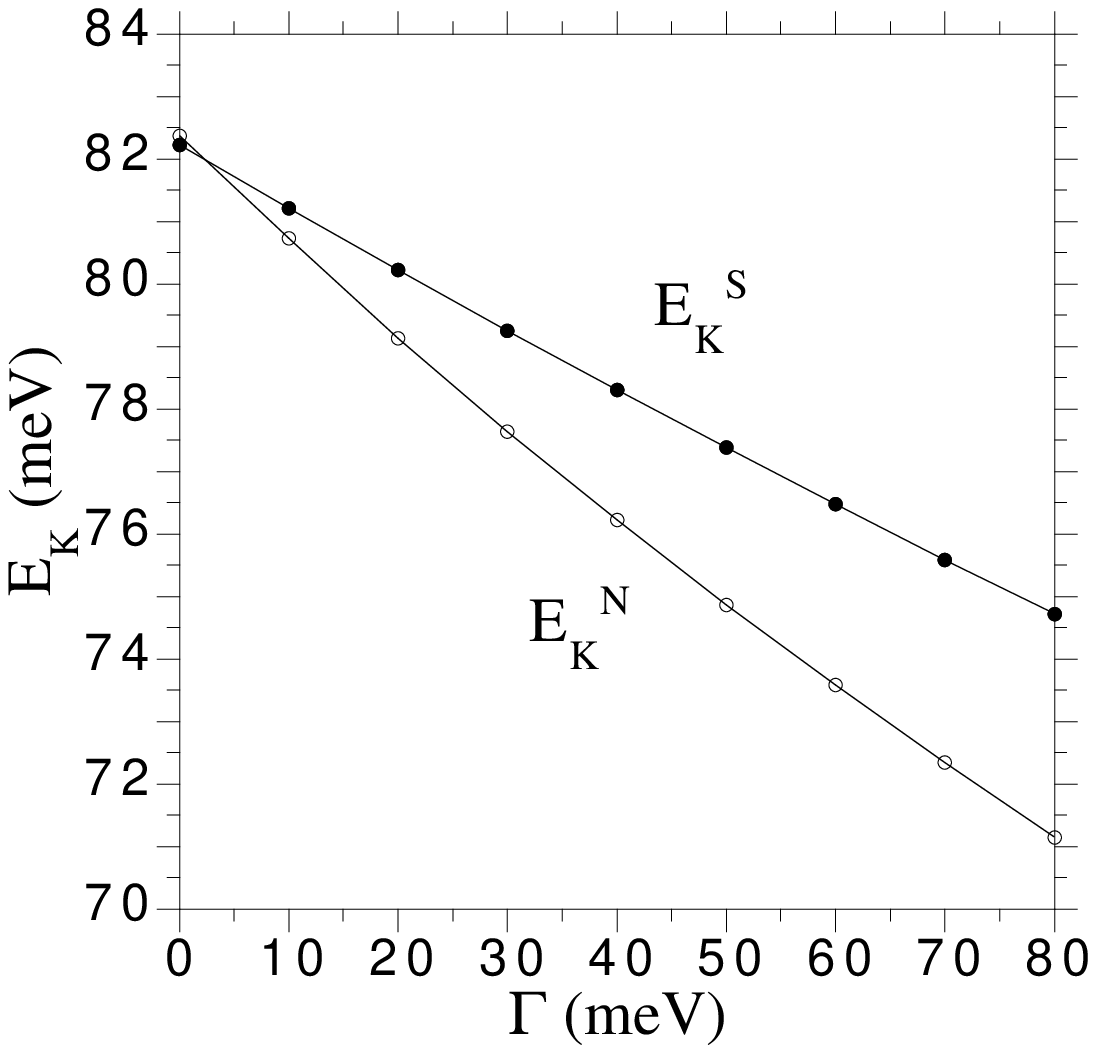}}
}
\vspace{0.3cm}
\caption{(a) $-\Delta E_K$ and (b) $E_{K}$ versus $\Gamma$.
$\alpha$=0.75, $\omega_0$=71meV,
$\Delta_{max}$=32meV, and $\omega_c$=500meV.}
\label{fig3}
\end{figure}

For illustrative purposes, we consider first the case with no anisotropy
in $\Gamma$.  In Fig.~3, we plot the sum rule violation versus
$\Gamma$ for a typical value of $\alpha$.  The variation with $\Gamma$ is
similar to Fig.~1a despite the presence of a
substantial $\alpha$ term. Thus for a ``pure" marginal
Fermi liquid ($\Gamma=0$), the sum rule violation is essentially zero.
The reason is that with $\Gamma=0$, the normal
state posseses quite sharp spectral peaks, and thus the change in $n_k$
when going into the superconducting state is reduced.

To make quantitative comparisons to experiment, realistic values of
$\Gamma_k$ and $\alpha$ as a function of doping are needed.
We can obtain them from the optics data.  In Fig.~4a,
we plot $1/\tau(\omega)$ for four Bi2212 samples in the
superconducting state extracted from reflectivity data\cite{PUCH}.  
The linear high frequency behavior is of the form $a + b\omega$.  Let us
relate these parameters to $\Gamma_k$ and $\alpha$.
The $\alpha$ term is easy to obtain, since it is $k$ independent.
At T=0, $1/\tau$ is an average of -2Im$\Sigma$ over a frequency range of
%MRN - added reference
0 to $\omega$ \cite{PUCH}.  Since the $\alpha$ term is linear in $\omega$, then
$\alpha=b$.

The $\Gamma_k$ term is a different story.  If it were isotropic, then
$\Gamma = a/2$.  For the anisotropic case, these two quantities are related
by a Fermi surface integral.  We can do this analytically by replacing the 
anisotropy term by $\cos^2(2\phi)$, where $\phi$ is the Fermi surface angle 
(the node is at $\phi=\pi/4$).  We find that the Fermi velocity along the 
Fermi surface
can also be fit to the same anisotropic form.  The resulting transport
%MRN -added reference
integral is \cite{MAHAN}
$2\tau(0) = [\int d\phi v(\phi)/\Gamma(\phi)]/\int d\phi v(\phi)$
where $v(\phi)=v_N[1 + v_R \cos^2(2\phi)]$ ($v$ is the modulus of the velocity)
and $\Gamma(\phi)=\Gamma_N[1 + c_R \cos^2(2\phi)]$.
From ARPES\cite{JS1} $c_R=3$, and from the tight binding fit,
$v_R=-0.72$.  Solving, we find that $\Gamma_N^{-1} = 3.4\tau(0)$.

For the other parameters, we note that the deviation of
$1/\tau(\omega)$ from linearity sets in at an energy $\Delta_{max}+\omega_0$,
where $\omega_0$ is the single particle scattering rate gap.
%MRN - change and addition
This is easily shown from the Kubo bubble by dressing one of the two lines.
From ARPES and tunneling,
$\omega_0 = \Delta_{max} + \omega_{res}$, where $\omega_{res}$ is the 
energy separation of the peak and the dip\cite{ND,JC99,JOHNZ}.
This is found to vary with doping as 5$T_c$\cite{JOHNZ}, and we use this to 
extract $\Delta_{max}$ from the optics scattering rate gap.
The resulting $\Delta_{max}$ values are
consistent with ARPES\cite{JC99} and tunneling\cite{JOHNZ} measurements.

%MRN - included experimental results, added about 1/tau offset
\begin{figure}
\centerline{
\epsfxsize=0.24\textwidth{\epsfbox{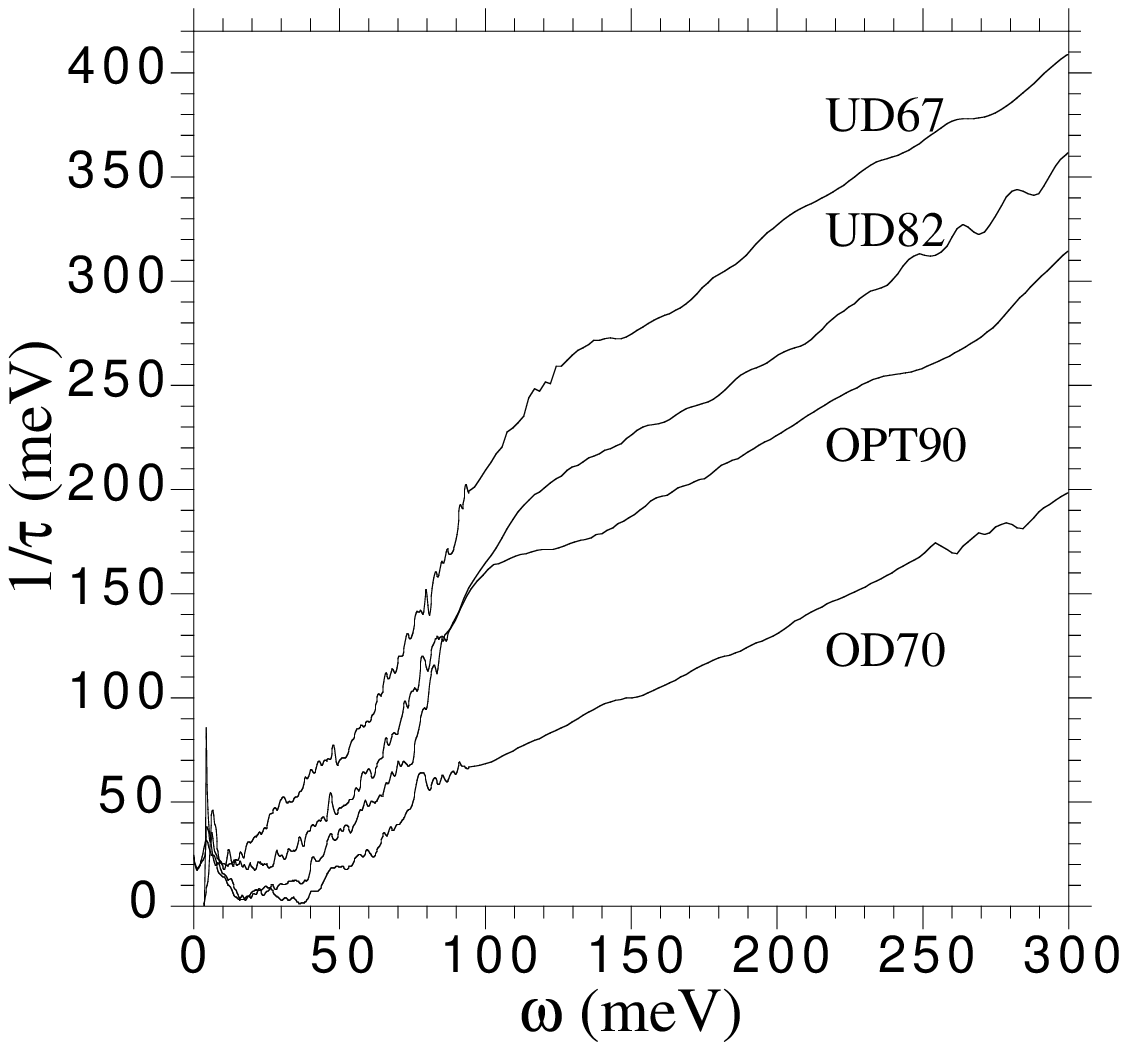}}
\epsfxsize=0.25\textwidth{\epsfbox{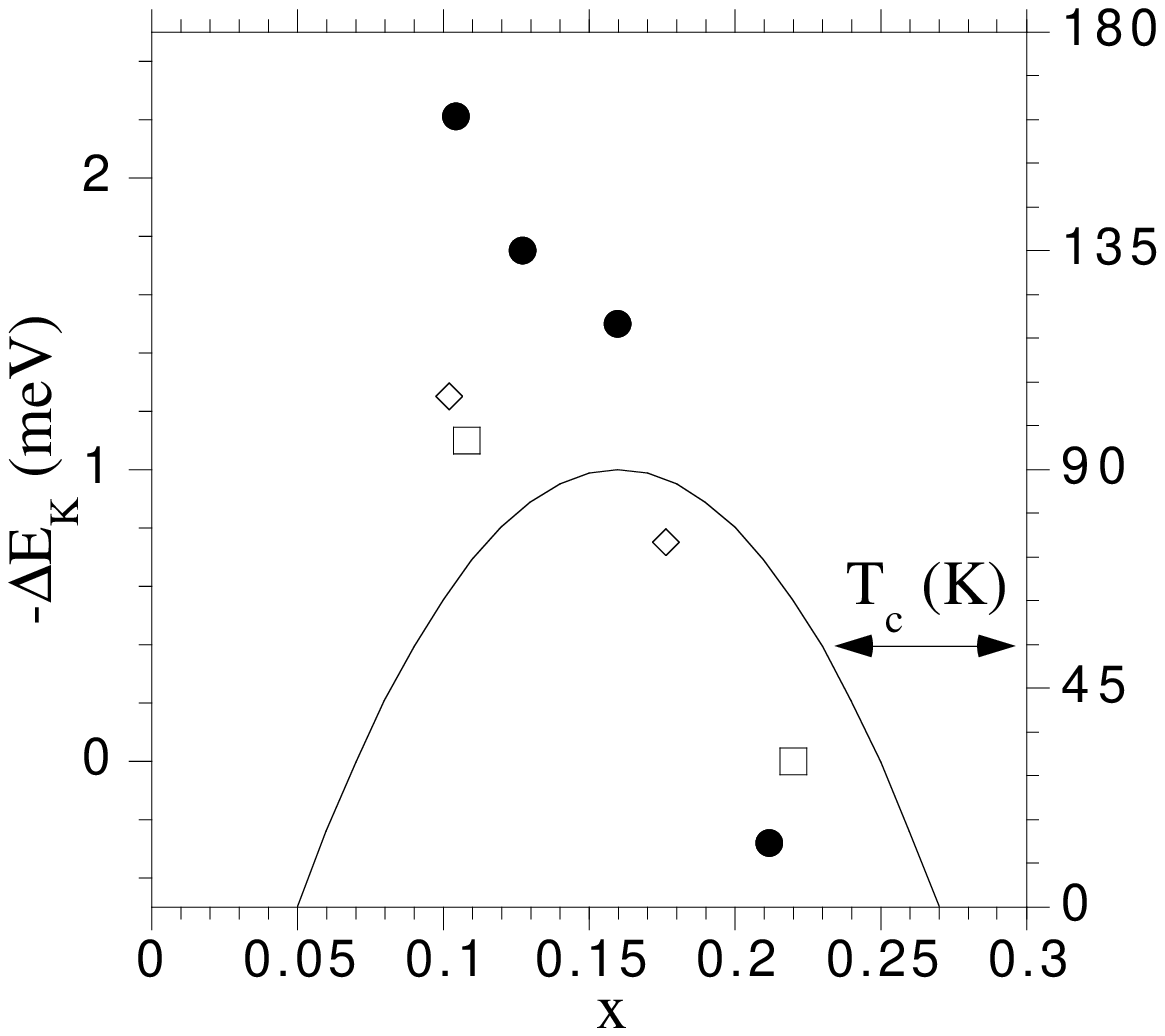}}
}
\vspace{0.3cm}
\caption{(a) $1/\tau(\omega)$ versus $\omega$ for various Bi2212 samples
from Ref.~\protect\onlinecite{PUCH} (OD overdoped, OPT optimal doped,
UD underdoped).  (b) Calculated sum rule violation
($-\Delta E_K$) versus doping, $x$.
The curve is $T_c$.  The parameters (meV)
extracted from (a) are $\Gamma_N$ (1, 22, 27, 37), $\alpha$ (.65, .75, .88,
.98), $\omega_0$ (54, 71, 76, 83), and $\Delta_{max}$ (24, 32, 41, 54) for
OD70, OPT90, UD82, and UD67 respectively.  Also shown in (b) are the 
experimental results (open squares from Ref.~\protect\onlinecite{NICOLE}, open 
diamonds from Ref.~\protect\onlinecite{VDM}).  The theoretical doping trend 
in (b) is due to the increasing offset in $1/\tau$ seen in (a).}
\label{fig4}
\end{figure}

%MRN - some changes in this paragraph to better reflect experiment
These values are used to determine the sum rule violation versus doping,
shown in Fig.~4b.  We find no sum rule
violation for the overdoped sample.  This is the expected BCS like behavior,
and is consistent with the experimental result on an overdoped
film\cite{NICOLE}.
For the other samples, we find a sum rule violation which increases
from 1.5 to 2.2 meV as the doping decreases.  The doping trend is consistent
with the reported experimental results (also shown in Fig.~4b), although the
values are perhaps too large by
a factor of two.  This may be due to the approximation of using a hard
infrared cut-off on the scattering rate in the superconducting state.  Still,
given the simplicity of our model, and the substantial experimental error
bars, the agreement with experiment is surprisingly good.  The doping trend
in our model is due to the increase in $\Gamma_N$ with underdoping.  We also
note that the kinetic energy change is about twice $-\Delta E_{K}$.

As for where the extra condensate weight is coming from, we note that the 
the experimental optical integrals balance at an energy cut-off of about
2 eV\cite{NICOLE}.  This value is comparable to the Mott gap of the insulator,
so we speculate that the extra weight comes from the upper and lower Hubbard
bands. This would be in accord with the more delocalized nature of the electrons
in the superconducting state.

In conclusion, using a simple model for the frequency dependent scattering
rate, we can understand recently reported results for the sum rule
violation for the in-plane conductivity.  The effect is due to the formation
of quasiparticles in the superconducting state, and confirms earlier 
speculations by Anderson\cite{PHIL}.  As the doping increases into the overdoped
region, we find the sum rule violation goes away, consistent with the more
Fermi liquid like nature of the normal state.

We acknowledge discussions with N. Bontemps, A.
Santander-Syro, D. van der Marel, D. Basov, T. Timusk,
and A. Georges.
This work was supported by the U.S. Dept.~of Energy, Office of Science, under
contract W-31-109-ENG-38,
the Aspen Center for Physics, and the CNRS (France).  We would like to thank
Prof. Bontemps and the ESPCI for their hospitality while this work was in
progress, Prof. Timusk for providing his data,
and Profs. Bontemps and van der Marel for use of their
unpublished work.


\begin{references}
\bibitem{LOR}
In calculations, we assumed a Lorentzian broadening of the poles with a half
width 5 meV.
\end{references}
\end{document}